\documentclass[fleqn,10pt]{wlscirep}
\usepackage[utf8]{inputenc}
\usepackage[T1]{fontenc}
\usepackage{upgreek}
\usepackage[makeroom]{cancel}
\usepackage{siunitx}
\usepackage{amsmath}

\title{S-polarized light-sheets improve resolution and light-efficiency in oblique plane microscopy}

\author[1]{Jon-Richard Sommernes}
\author[2]{Alfred Millett-Sikking}
\author[1,*]{Florian Ströhl}
\affil[1]{Department of Physics and Technology, UiT The Arctic University of Norway, Tromsø, Norway}
\affil[2]{Calico Life Sciences LLC, South San Francisco, CA, USA}

\affil[*]{florian.strohl@uit.no}

\begin{abstract}
Oblique plane microscopy (OPM) offers 3D optically sectioned imaging with high spatial- and temporal-resolution while enabling conventional sample mounting. The technique uses a concatenation of three microscopes, two for remote focusing and a tilted tertiary microscope, often including an immersion objective, to image an oblique sample plane. This design induces Fresnel reflections and a reduced effective aperture, thus impacting the resolution and light efficiency of the system. Using vectorial diffraction simulations, the system performance was characterized based on illumination angle and polarization, signal to noise ratio, and refractive index of the tertiary objective immersion. We show that for samples with high fluorescent anisotropy, s-polarized light-sheets yield higher average resolution for all system configurations, as well as higher light-efficiency. We also provide a tool for performance characterization of arbitrary light-sheet imaging systems.
\end{abstract}
\begin{document}

\flushbottom
\maketitle
\thispagestyle{empty}

\section*{Introduction}

Volumetric imaging is an important tool for studying biological phenomena, and has thus given rise to a broad set of techniques with various benefits and drawbacks\cite{fischer2011microscopy}. One of these techniques is light-sheet fluorescent microscopy (LSFM)\cite{mappes2012invention,voie1993orthogonal,stelzer1994fundamental,stelzer1995new,chen2014lattice,huisken2009selective,santi2011light}. LSFM offers a combination of high spatial- and temporal-resolution, optical sectioning, and low phototoxicity, which makes the technique highly suitable for live cell imaging. LSFM works by having two orthogonally oriented objectives with a shared foci. One objective illuminates the focus plane of the second objective with a light-sheet during imaging. This technique has given rise to several designs over the last decades \cite{voie1993orthogonal,stelzer1994fundamental,stelzer1995new,chen2014lattice,huisken2009selective}. However, because the objectives are placed orthogonal to each other while maintaining a shared foci, the combined Numerical Aperture (NA) of the two objectives is spatially limited. Additionally, the physical design of the two-objective setup generally prevents the system from utilizing conventional sample mounting. This makes the technique challenging for certain sample types like slides, dishes, and multiwell plates.

\begin{figure}[hbt!]
    \centering
    \fbox{\includegraphics[width=0.9\linewidth]{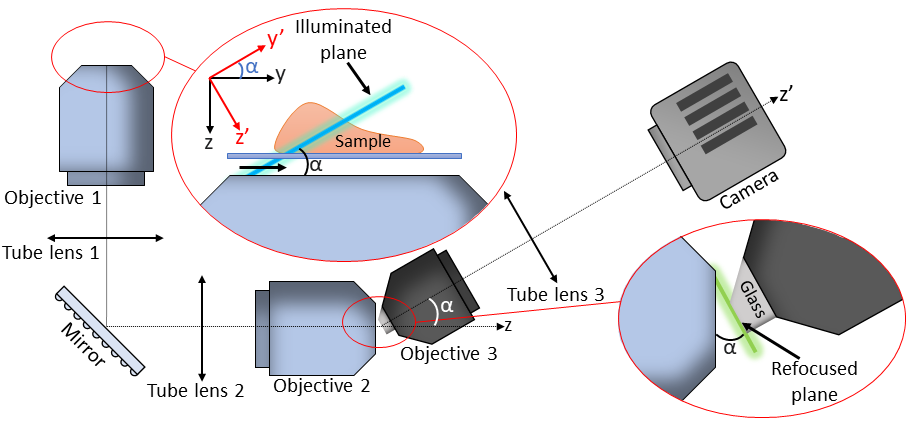}}
    \caption{Schematic setup of an OPM system. O and T represent objectives and tube lenses respectively. The lens train from O1 to O2 forms a perfect 3D imaging system, and O3 and T3 forms a tertiary tilted microscope. The sample is placed above O1, and illuminated by a light-sheet (blue) inclined by a tilt $\upalpha$. The illuminated area is relayed onto the focus of O3, and re-imaged onto the camera.}
    \label{fig:OPM_setup}
\end{figure}

\textbf{Oblique Plane Microscopy} (OPM)\cite{dunsby2008optically,bouchard2015swept,kumar2018integrated,sparks2020dual,glaser2022hybrid,yang2022daxi,sapoznik2020versatile} is a technique that uses the same objective lens for both illumination and imaging, thus overcoming the main limitations of LSFM. A schematic OPM setup is shown in Figure \ref{fig:OPM_setup}. In this setup, the light-sheet is formed by illuminating the pupil of the imaging objective off center. The center point of the pupil illumination will determine the inclination of the light-sheet, while the extent of the illumination will determine the thickness, width, and length of the focused sheet. This will generate a light-sheet at an oblique angle to the focal plane of O1. Where O represents a microscope objective, and T represents a tube lens.

Due to the oblique illumination, most of the illuminated area will be out of the traditional focus plane. To get this plane in focus, the sample volume needs to be relayed to a remote image space using a perfect 3D imaging system\cite{botcherby2008optical}, seen in Figure \ref{fig:OPM_setup} as the lens train from O1 to O2. To avoid aberrations and distortions, the magnification of a perfect imaging system should be close to unity\cite{mohanan2022sensitivity}. The remote image space is then re-imaged by a third microscope, seen in Figure \ref{fig:OPM_setup} as O3 and T3. This microscope is oriented so the remote image plane is perfectly in the traditional focus plane of O3. 

Here, we assume the effective pupil function of this system will then be limited by the overlap of the O2 and O3 pupil. This effective pupil will limit the achievable resolution of the system. To maximize the effective pupil, a high NA O3 can be used. However, due to the tilted geometry of the tertiary microscope, an immersion medium is needed to fit a high NA O3. A specialized glass immersion objective (P/N: AMS-AGY), often called “snouty”\cite{alfred_millett_sikking_2019}, has been created for this purpose, although water immersion objectives using glass cover slips have also been used\cite{yang2019epi}. This change in RI from O2 to O3 can make the light transmission highly polarization dependent. This dependency will be especially noticeable for samples with high fluorescent anisotropy\cite{lakowicz2006principles}. 

Fluorescent anisotropy is a measure of how much of the excitation polarization is preserved in the emission. In this paper we model fluorophores as ensembles of rotating dipoles. If a dipole has low rotational diffusion, e.g. from a chemical bond or high mass, the anisotropy of the dipole is increased. In this paper, we explore the extreme case of low rotational diffusion, corresponding to maximum retention of the excitation polarization.

\section*{Methods}

In order to investigate the impact of polarization on the system, a simulation software was developed to model the vectorial diffraction of the system as outlined by Kim, et al\cite{kim2018calculation}. In the simulations, an emission source was modelled as a point-like dipole and placed in the focal point of O1. This was done to immitate the emission pattern of a small fluorophore. The electric field of the dipole emission was then characterized at discreet points within the collection cone of O1. In order to trace the fields through the optical system, each optical component was modelled as a 3$\times$3 Jones matrix. To find the complete transformation of the system, the matrices were then multiplied to find a system transformation. For our case, the system model can be described as:

\begin{align}
    \textbf{T}_{sys}(\theta_1,\phi,\alpha) = &\textbf{R}_{\textbf{z}}^{-1}(\phi_{\alpha}) \textbf{M}_3(\theta_3) \textbf{R}_\textbf{y}(\theta_3) \textbf{F}_\textbf{T} \textbf{R}_\textbf{y}(-\theta_2) \textbf{R}_\textbf{z}(\phi_{\alpha}) \textbf{R}_\textbf{x}(\alpha) \textbf{R}_\textbf{z}^{-1}(\phi) \overline{\textbf{M}}_2(\theta_2) \textbf{M}_1(\theta_1) \textbf{R}_\textbf{z}(\phi) \underline{E}_0 \\
    \textbf{M}_i(\theta_i) = &\textbf{T}_i(\theta_i') \textbf{O}_i(\theta_i) \boldsymbol{\Gamma}_i(\theta_i) A_i(\theta_i) \text{\quad , \quad}
    \overline{\textbf{M}}_i(\theta_i) = \textbf{O}_i(\theta_i) \textbf{T}_i(\theta_i') \boldsymbol{\Gamma}_i(\theta_i) A_i(\theta_i)
\end{align}

where $\textbf{R}_\textbf{i}$ denotes the rotation matrix around axis \textbf{i}, \textbf{O} and \textbf{T} are both lens matrices but separated to distinguish between tube lenses and objectives, $\boldsymbol{\Gamma}_i$ is the light transmission of the objective at $\theta_i$, $A_i$ is the apodization of the microscope at $\theta_i$, $\textbf{F}_\textbf{T}$ is the Fresnel transmission of the refractive index (RI) change, and $\underline{E}_0$ is the electric field of the dipole emission. The Fresnel transmission is only applied if the transmission to a new immersion medium is not altered by a anti-reflection coating. A full description of the matrices can be found in the supplementary information.

\begin{figure}[ht!]
    \centering
    \fbox{\includegraphics[width=0.9\linewidth]{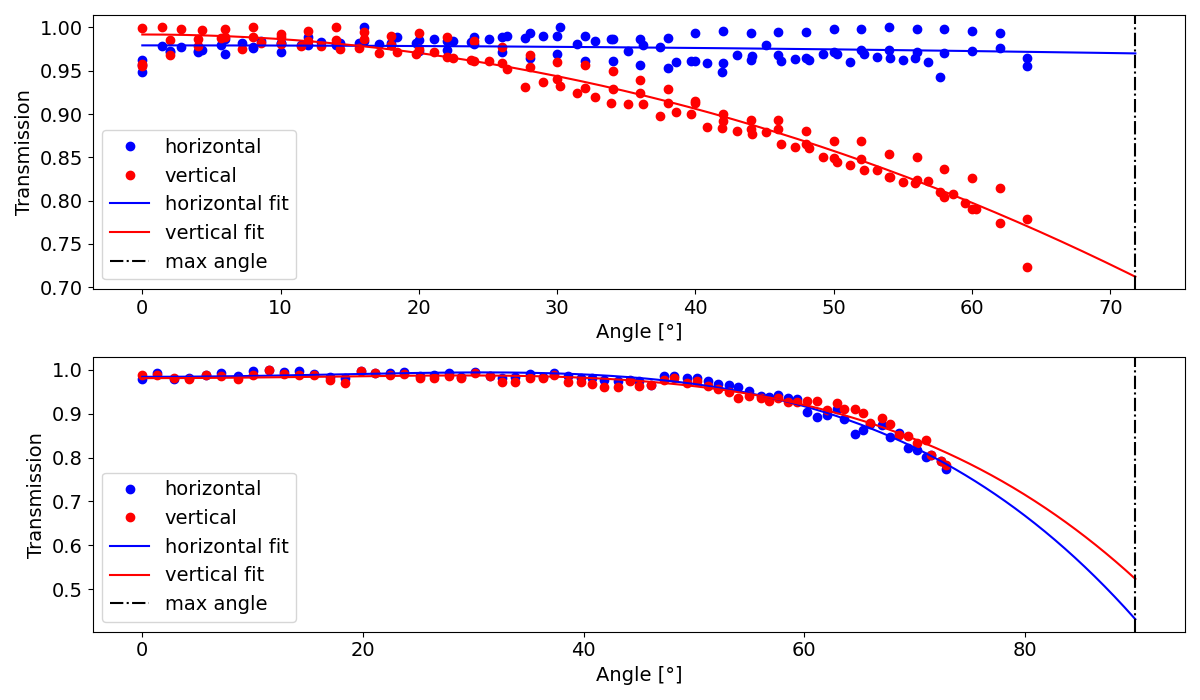}}
    \caption{The top panel shows the light transmission for a 0.95NA dry objective (CFI Plan Apochromat Lambda D 40X), while the bottom panel shows the light transmission for a 1NA glass immersion objective (AMS-AGY v1.0). The blue and red lines indicate p- and s-polarized light respectively. The transmission of the objectives was normalized to the maximum value and curve fitted using 5th degree polynomials.}
    \label{fig:objective_transmission}
    \end{figure}

For the glass immersion objective (AMS-AGY v1.0), and the dry objective used for O2, the transmission was measured experimentally using a single-axis stage and a power meter. A laser, beam expander, tube lens, polarizer, and half-wave plate was built into a single unit and mounted on a translation stage. The laser beam was expanded to fill a tube lens, then focused onto the back focal plane (BFP) of the objective. By moving the focus of the beam radially from the center of the BFP, the output angle of the beam was changed. The power of the beam was then measured at a range of angles for both p- and s-polarized light. The results were then curve fitted to 5th degree polynomials. This polynomial can be assumed to be a taylor approximation of the true transmission function. The function was then used to make a transmission mask for the corresponding objective. The data used for this mask is found in Figure \ref{fig:objective_transmission}.

Further, for every microscope the lenses are assumed to share an identical BFP. Assuming all lenses follow the abbe sine condition, the angles of the simulated rays can be related using\cite{gu2000advanced}:

\begin{equation}
    \theta_i' = \arcsin\left(\frac{\text{NA}'}{\text{NA}}n\sin(\theta_i)\right)
\end{equation}

where $\text{NA}$ is the Numerical Aperture of the objective, $\text{NA}'$ is the Numerical Aperture of the tube lens, $\theta_i$ is the angles of rays at the objective, $\theta_i'$ is the anlges of the rays at the tube lens, and $n$ is the refractive index of the objective immersion medium.

For two lenses sharing an image space, the lenses are assumed to share an identical focal point. To get the input angles of the collimating lens, the output angles of the focusing lens is altered using either Snell's law for a immersion medium change, or a change in the optical axis for the transition from O2 to O3. 

To ensure the validity of the model, some assumptions were made for the system: the dipole emitter is located perfectly in the focal plane of the first objective, the lenses are rotationally symmetric and centered on the optical axis, and there are no spherical or chromatic aberrations.

When the electric field has been traced through the system, we end up with the final electric field converging towards the focal point. This electric field can then be evaluated using a version of the Debye integral\cite{debye1909behavior}. Owing to the computational inefficiency of numerical integration, the integral was evaluated using a Fourier transform version of the Debye integral. The version of the Debye integral used here was based on the method outlined by Leutenegger et al.\cite{leutenegger2006fast}, then modified and implemented by James Manton\cite{Manton2022debye}. To get sufficient sampling of our signal, the pixel size, $v$, of our camera needs to be sufficiently small. To fulfill this condition, Nyquist sampling is normally sufficient. For an in depth derivation of this condition, see the supplementary information.




Having set a suitable pixel size, we need to find a sampling volume that is large enough to avoid significant PSF truncation. As the maximum extent of the PSF is in the axial direction, we can find the sampling volume by the axial extent of the PSF. Assuming the PSF can be modeled as an Airy disc, the PSF will extent until the signal is below the noise floor. Although the mathematical formulation of the Airy disc can be used to determine this, it is more practical to test different sampling volumes until the results converge. For our simulations, a sampling size of 240$\upmu$m in each axis was found to be mostly sufficient.





Now we have set constrains on the pixel size and count. The remaining simulation parameters are only constrained by the specific setup, and can thus be freely chosen without impacting the validity of the simulation. The remaining simulation parameters are: the light-sheet inclination and polarization, the fluorescent anisotropy and brightness of the sample, the excitation and emission wavelength, the lenses used, and the size of the dipole ensemble used for the PSF calculation. The light-sheet thickness is determined by its opening angle. As we are, in this paper, interested in the limits of each system configuration, we assume the light-sheet is made using the maximum available aperture. The parameters used for the simulations will be given in the results section, alongside the corresponding result.

To predict the Point Spread Function (PSF) that will be read from a camera, the noise-less PSF was normalized to achieve the desired photon count. Photon shot noise was then added by passing the voxels through a Poisson distribution, followed by the addition of Gaussian noise to simulate camera readout noise. Once noise was added, a bias offset was included and the voxels were discretized.

To determine the Optical Transfer Function (OTF), the PSF was Fourier transformed. The OTF contains the resolvable frequencies of an image, and by identifying the maximum extent of the OTF in each axis, the corresponding resolution limit can be defined. Assuming a PSF with shot noise, Gaussian readout noise, and a bias offset, the background of the OTF can be approximated using:



\begin{equation}
    |\hat{\eta}_X| = \sqrt{DC_{\hat{X}} + \sigma_{RMS}^2 - b}
\end{equation}

where $|\hat{\eta}_X|$ is the average background power of the Fourier transform, $DC_{\hat{X}}$ is the DC component of the Fourier transform, $\sigma_{RMS}$ is the RMS of the readout noise, and $b$ is the bias offset. For a rigorous derivation of this background, see the supplementary information.

Subtracting this background from the OTF, the cutoff frequency is given by the first zero crossing of the modified OTF. The resolution limit is then found using the cutoff frequency and the base frequency of the Fourier transform. However, in $\hat{z}$-direction the OTF will have two different cutoff frequencies, one along the $\hat{z}$-axis and a second for the maximum extent in the $\hat{z}$ direction. The cutoff along the $\hat{z}$-axis will give a sectioning thickness, while the maximum extent will give the axial resolution limit. Then, assuming the PSF is an ellipsoid, we can find the in-focus area and volume of the PSF. For this paper, we will use the sectioning thickness to calculate the PSF volume.

\section*{Results}

\begin{figure}[hbt!]
  \centering
  \fbox{\includegraphics[width=0.9\linewidth]{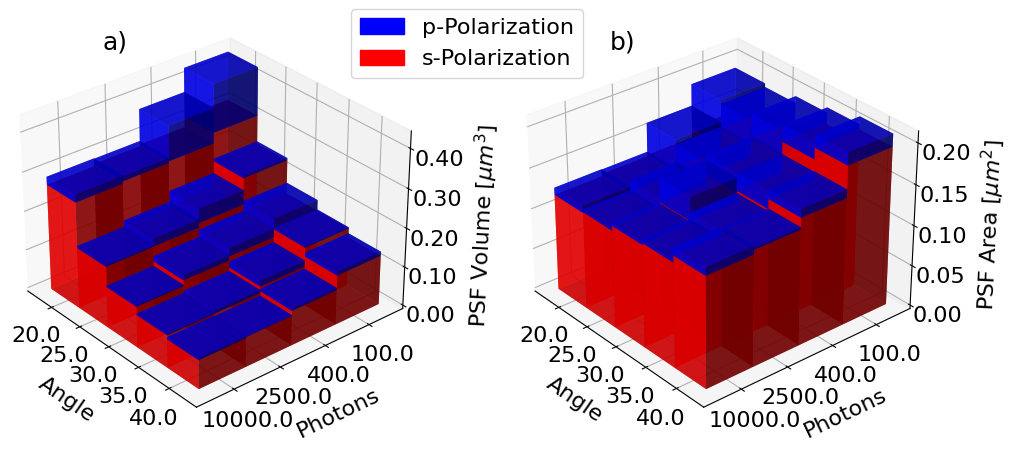}}
  \caption{Panels a) and b) shows the PSF volume and area respectively for all system configurations using an AMS-AGY v1.0 O3. The blue and red bars represent p- and s-polarized light-sheets respectively. The left axis represents a range of light-sheet inclinations (angle from focus plane to light-sheet plane), while the right axis represents the photon count at the brightest pixel. The PSF area and volume were calculated assuming an elipsoidal PSF. All results are derived from ten simulations with identical system configuration and an ensemble of 100 dipoles.}
  \label{fig:resolution_plot}
\end{figure}

\begin{table}[hbt!]
  \centering
  \caption{\bf Resolution limits, sectioning, and light-sheet length for all tested system configurations}
  \begin{tabular}{cccccccc}
    \hline
    & \multicolumn{2}{c}{x {[}nm{]}} & \multicolumn{2}{c}{y {[}nm{]}} & z {[}nm{]} & S {[}nm{]} & $\mathrm{L}_{ls}$ [\SI{}{\um}] \\
    $\upalpha$ & p & s & p & s & p/s & p/s &  \\
    \hline
    20$^\circ$ & 188 & 185 & 216 & 205 & 571 & 1714 & 57.3  \\
    25$^\circ$ & 185 & 182 & 212 & 209 & 457 & 980  & 14.4  \\
    30$^\circ$ & 185 & 187 & 222 & 214 & 403 & 623  & 6.5 \\
    35$^\circ$ & 189 & 186 & 223 & 219 & 343 & 490  & 3.7 \\
    40$^\circ$ & 198 & 188 & 235 & 231 & 286 & 403  & 2.4 \\
    \hline
  \end{tabular}
  \label{tab:resolution}
\end{table}

\begin{figure}[htb!]
  \centering
  \fbox{\includegraphics[width=0.9\linewidth]{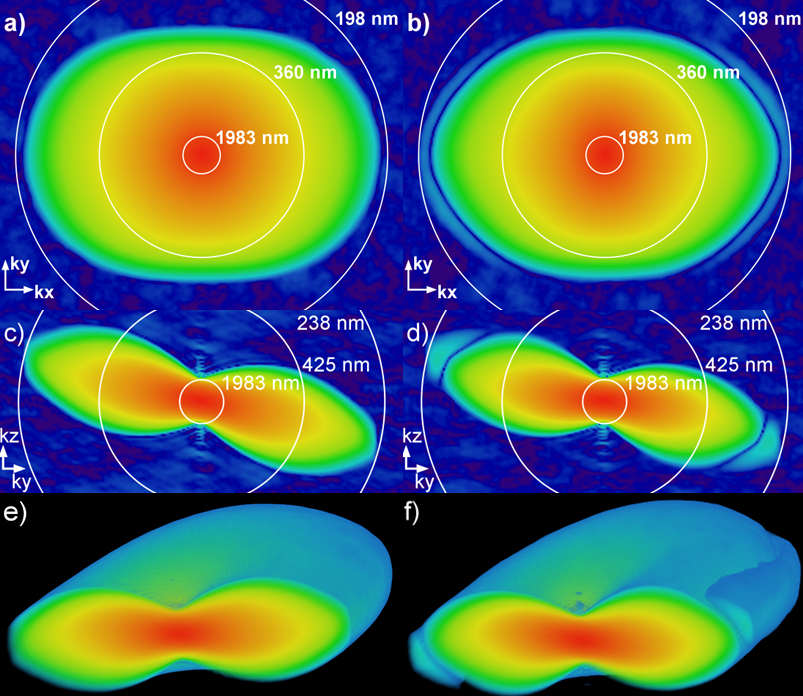}}
  \caption{The OTF of a system configuration using 30° light-sheet inclination, 1.25NA water immersion O1, 0.95NA dry O2, 1.0 NA glass immersion O3, and SNR=100. Panels a), c), and e) show the OTF for an s-polarized light-sheet, while panels b), d), and f) show the OTF for a p-polarized light-sheet. Panels a) and b) show the XY cross section of the OTF, c) and d) show the YZ cross section, and e) and f) show a 3D rendering of the OTF.}
  \label{fig:OTF_CS}
\end{figure}

\begin{figure}[hbt!]
  \centering
  \fbox{\includegraphics[width=0.9\linewidth]{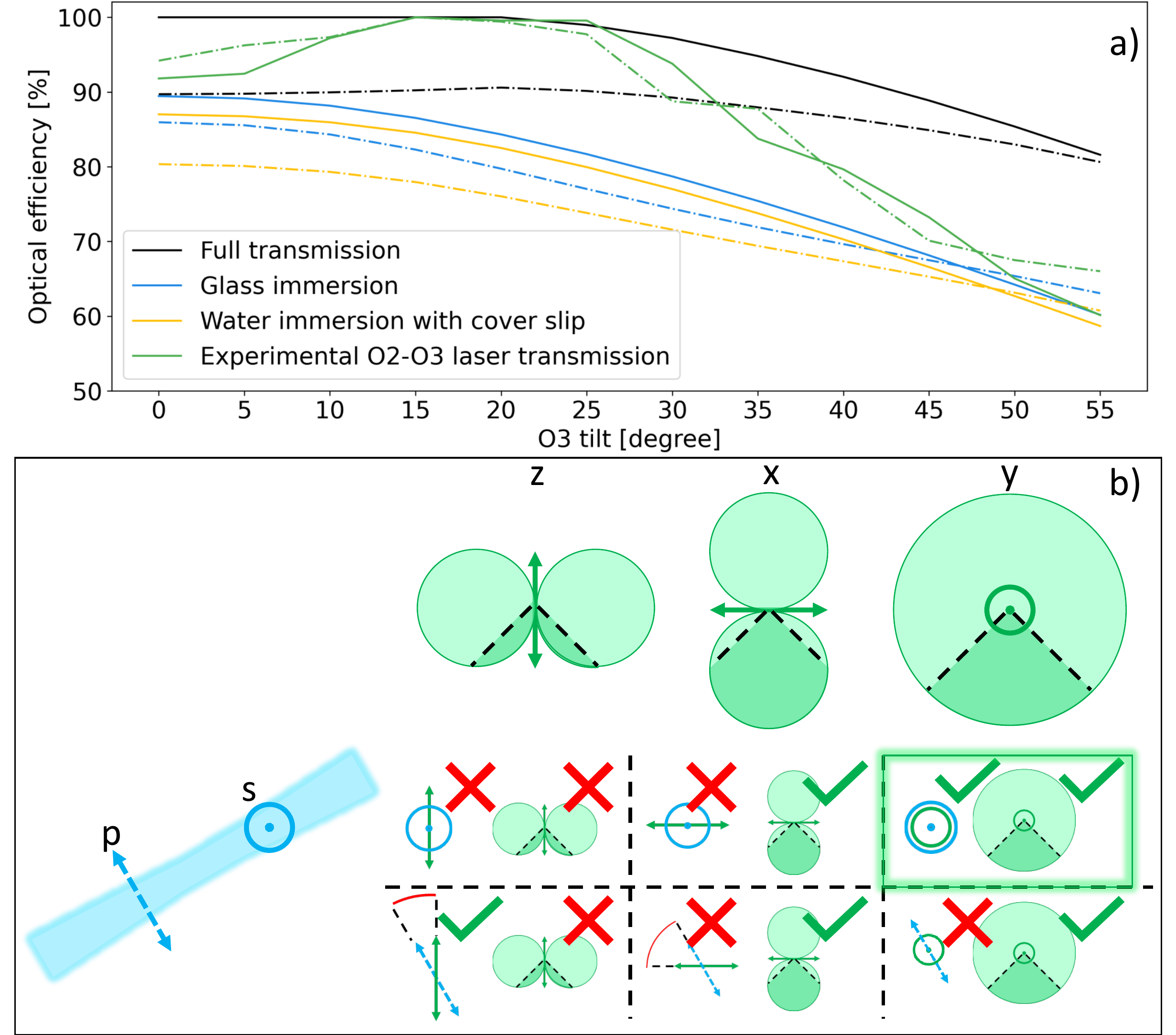}}
  \caption{Optical efficiency of OPM. a) shows the amount of light emitted by the fluorophore that is transmitted through the system, normalized to the maximum transmission. The black lines represent a 1NA dry O3 (physically unrealized, but simulated to remove Fresnel reflections). The blue lines represent an AMS-AGY v1.0 style O3. The yellow lines represent water immersion O3 using a 1.5 RI cover slip. The green lines show the transmission of a collimated laser transmitted through O2 and O3, measured experimentally. The dotted lines represent p-polarized input light, and solid lines represent s-polarized light. b) shows pictogram representing the excitation and emission of the dipole emitters. The light-sheet (blue) can be p-polarized or s-polarized, indicated by the axes drawn on the bottom left and top right of the light-sheet respectively. The green shapes above the matrix represent the emission pattern of a dipole oriented along the z-, x- and y-axis going from left to right. The acceptance cone of the objective is indicated by the dark-shaded area. In the matrix, each cell consist of two elements. On the left of the cell is the excitation efficiency of the light-sheet on the dipole, and on the right is the collection efficiency of the dipole emission. A red cross represent low efficiency, and a green check mark represent high efficiency.}
  \label{fig:light_efficiency}
\end{figure}

Figure \ref{fig:resolution_plot} shows the PSF volume and is using a 1.35NA silicone immersion O1, 0.95NA dry O2, and a 1NA glass immersion O3. The results use an ensemble of 100 dipoles to make the effective PSF. To minimize the effects of the random noise added to the PSF, the OTF was calculated for ten noise samples drawn from the same distribution. The resolution limits were then found for all ten noise samples, and averaged to find a single value. The axis specific resolution limits, as well as the sectioning thickness and length of the light-sheet, are given in Table \ref{tab:resolution}. This table shows the resolution limits for the highest simulated SNR value (using $10^4$ photons at the brightest pixel). The results show an increase in PSF area and a decrease in PSF volume when increasing the light-sheet inclination. The results also show an effect on both PSF area and volume with respect to photon count. Depending on the system configuration and SNR, s-polarized light can give anything from 2\% to 30\% decreased PSF area and volume.

Figure \ref{fig:OTF_CS} shows an OTF for a system configuration using a 30° light-sheet inclination, 1.25NA water immersion O1, 0.95NA dry O2, AMS-AGY v1.0 O3, and SNR=100. The left panels (a, c, and e) show the OTF for an s-polarized light-sheet, while the right panels (b, d, and f) show the OTF for a p-polarized light-sheet. From the images, we can see a ring feature in the OTF for a p-polarized light-sheet that is not present in the s-polarized case.

Figure \ref{fig:light_efficiency}, panel a) shows the optical efficiency for different system configurations. All system configurations use a 1.35NA silicone immersion O1, 0.95NA dry O2, and a 1NA O3. The black lines represent a system with a dry O3. Albeit a 1.0 NA dry objective does not exist, it is included to show the theoretically optimal case. The yellow lines represent a water immersion O3 with a 1.5 RI glass coverslip to hold in the water\cite{yang2019epi}. The blue lines represent an AMS-AGY v1.0 O3. From this, we see that an AMS-AGY v1.0 objective gives higher optical efficiency than its water immersion counterpart thanks to its anti-reflection coating. We also find that s-polarized light-sheets have a higher optical efficiency than p-polarized light for most light-sheet inclinations. Only at extreme tilt angles do p-polarized light-sheets outperform s-polarized illumination in terms of light-efficiency.

Panel b) in Figure \ref{fig:light_efficiency} shows a pictogram describing the light-sheet excitation and dipole emission. The light-sheet (blue) can be p-polarized or s-polarized, indicated by the axes drawn on the light-sheet. The green shapes above the matrix represent the emission pattern of a dipole in various orientations. The excitation efficiency of the light-sheet on the dipole is given as a inner product of the polarization of the dipole and light-sheet. The collection efficiency of the dipole emission is given as the overlap between the emission spectrum and the acceptance cone of the objective lens. For a system configuration to be light efficient, both the excitation and collection efficiency must be good. As shown in the figure, this is only achieved for s-polarized light-sheets, with a dipole oriented along the y-axis.

\section*{Discussion}

In the results shown in Figure \ref{fig:resolution_plot}, we see an increase in PSF area with increased angle. This is expected, as the overlap of the O2 and O3 acceptance cone will decrease with increased angle. This results in a decrease in the effective aperture of the system, thus increasing the PSF area. The PSF volume, however, is seen to decrease with increased angle. As the light-sheet inclination increases, the aperture available for making the light-sheet increases, thinning the light-sheet, and thus decreasing the effective PSF volume. However, a system can use a fixed light-sheet thickness for any inclination. In this case, the length of the light-sheet will be contant. This allows for imaging deeper into a sample, but at the cost of volumetric resolution. The light-sheet volume will then follow the trend of the simulated PSF area.


The effect of light-sheet polarization is seen in Figure \ref{fig:OTF_CS}. Here, we see that p-polarized light-sheets give rise to side lobes of low power in the OTF. These side lobes are likely a result of a sign change in the OTF and are prone to blend into the background more quickly. This effect is not present for s-polarized light-sheets, making the higher frequency components of the OTF more visible for s-polarized light-sheets. Also found in Figure \ref{fig:OTF_CS} are artifacts from the Fourier transform along the kz axis (seen in panels c) and d)). These artifacts are a result of PSF truncation, and can be minimized by increasing the sampling volume.

The results shown in Figure \ref{fig:resolution_plot} are simulated using an AMS-AGY v1.0 style O3. As seen in Figure \ref{fig:objective_transmission}, the light throughput of this objective is minimally affected by the polarization, which is not the case for a water immersion O3 with coverslip. Note that an anti-reflection coating could be applied to a coverslip to provide similar resilience to polarization and improved transmission performance. In this case, the light throughput will be highly dependent on the polarization of the light-sheet. This will likely alter the results shown in Figure \ref{fig:resolution_plot}. 

Further, the immersion medium of O3 will affect the light transmission of the system, as seen in Figure \ref{fig:light_efficiency}. Here we see that s-polarized light-sheets have a higher optical efficiency than p-polarized light-sheets. This effect is illustrated in panel b) of Figure \ref{fig:light_efficiency}. As an s-polarized light-sheet is the most effective at exciting a dipole oriented along the y-axis, the collection efficiency of the emission will be higher for s-polarized light-sheets. However, due to better transmission of p-polarized light after O1, as seen in Figure \ref{fig:objective_transmission}, this gap in optical efficiency might decrease for certain system configurations. This increased transmission is not, however, sufficient to overcome the increased collection efficiency of s-polarized light-sheets. This means that s-polarized light-sheets will almost always have a higher optical efficiency than p-polarized light-sheets. This effect has only been simulated for a non-scattering medium. In scattering samples, other effects might come into play. For instance, p-polarized light is more likely to be scattered in the plane of the light-sheet, thus reducing unwanted background light excitation in dense samples.

The results also show that a AMS-AGY v1.0 style O3 gives lower light-loss than the Fresnel reflections caused by a water immersion O3 using a glass coverslip. This can be explained by the anti-reflection coating on the AMS-AGY v1.0. Further, the light transmission of the AMS-AGY v1.0 has been measured experimentally, while the water immersion O3 has only been simulated using Fresnel transmissions. This means the transmission loss will likely be higher for the water immersion O3 than shown in Figure \ref{fig:light_efficiency}, which would further increase the gap in optical efficiency between the two objectives. If no anti-reflection coated coverslip is used with a water-immersion lens, we conclude that the AMS-AGY O3 is the better choice for most OPM systems

\section*{Conclusion}

In this paper we have found that s-polarized light-sheets offer a more stable OTF, resulting in up to 30\% reduction in PSF volume for samples with high fluorescent anisotropy. We have also shown s-polarized excitation light to give up to 10\% increased optical efficiency for certain system configurations. We have found that s-polarized light-sheets increase optical efficiency for inclinations up to approximately 45$^{\circ}$. This increase in both achievable resolution and optical efficiency makes s-polarized light-sheets the better choise for most OPM setups. 

We also provide a software tool for simulating any arbitrary light-sheet imaging system (see Additional information). This software will allow users to find the PSF, OTF, and light transmission of any light-sheet imaging system as given in this paper.

\bibliography{sample}

\section*{Author contributions statement}

F.S. and J.R.S. conceived the simulations, A.M.S. and J.R.S conducted physical measurements, and J.R.S. wrote the simulations and analysed the results. All authors reviewed the manuscript.  

\section*{Additional information}

\textbf{Data Availability} Simulation software including user interface is found at https://doi.org/10.18710/YAYNNL.

\noindent\textbf{Competing interests} Calico holds IP on the AMS-AGY objective. A.M.S. is an employee of Calico Life Sciences LLC, and one of the inventors of the AMS-AGY. The other authors declare that there are no competing interests.

\noindent\textbf{Funding} The Research Council of Norway (\#314546) and the Centre for Digital Life Norway.

The corresponding author is responsible for submitting a \href{http://www.nature.com/srep/policies/index.html#competing}{competing interests statement} on behalf of all authors of the paper. This statement must be included in the submitted article file.

\end{document}


\maketitle

\section*{Tracing matrices}

In all the simulations in this paper, we have used the Jones calculus to trace the electric field through the system. The Jones calculus is a matrix formalism that describes the polarization of light. For our case, we have modeled the system using the following matrices:

\begin{align}
    \textbf{T}_{sys}(\theta_1,\phi,\alpha) = &\textbf{R}_{\textbf{z}}^{-1}(\phi_{\alpha}) \textbf{M}_3(\theta_3) \textbf{R}_\textbf{y}(\theta_3) \textbf{F}_\textbf{T} \textbf{R}_\textbf{y}(-\theta_2) \textbf{R}_\textbf{z}(\phi_{\alpha}) \textbf{R}_\textbf{x}(\alpha) \textbf{R}_\textbf{z}^{-1}(\phi) \overline{\textbf{M}}_2(\theta_2) \textbf{M}_1(\theta_1) \textbf{R}_\textbf{z}(\phi) \\
    \textbf{M}_i(\theta_i) = &\textbf{T}_i(\theta_i') \textbf{O}_i(\theta_i) \Gamma_i(\theta_i) A_i(\theta_i) \text{\quad , \quad}
    \overline{\textbf{M}}_i(\theta_i) = \textbf{O}_i(\theta_i) \textbf{T}_i(\theta_i') \Gamma_i(\theta_i) A_i(\theta_i)
\end{align}

All the matrices, except $\textbf{F}_\textbf{T}$, are simply rotation matrices. These matrices will map the electric field into a new coordinate system, without altering the amplitude of the field. In this section, we will describe the matrices and their role in the system.

The first matrix applied to the electric field is $\textbf{R}_\textbf{z}(\phi)$, and is defined as:

\begin{equation}
    \textbf{R}_\textbf{z}(\phi) = \begin{bmatrix}
        \cos(\phi) & \sin(\phi) & 0 \\
        -\sin(\phi) & \cos(\phi) & 0 \\
        0 & 0 & 1
    \end{bmatrix}
\end{equation}

where $\phi$ is the angle of rotation around the z-axis. This matrix is used to map the field from Cartesian coordinates into a the meridional and sagittal planes. This way, the electric field can eaily be traces through a lens using $\textbf{L}(\theta)$, which is defined as:

\begin{equation}
    \textbf{L}(\theta) = \begin{bmatrix}
        \cos(\theta) & 0 & \sin(\theta) \\
        0 & 1 & 0 \\
        -\sin(\theta) & 0 & \cos(\theta)
    \end{bmatrix}
\end{equation}

where $\theta$ is the angle of the ray relative to the optical axis. Both $\textbf{O}_(\theta)$ and $\textbf{T}_(\theta)$ are identical to $\textbf{L}(\theta)$, but distinguished for clarity to separate objectives and tube lenses. This matrix is used to simulate a lens, which is done by either collimating or focusing the electric field. The matrix $\boldsymbol{\Gamma}(\theta)$ is used to map the transmission of the lens, and is defined as:

\begin{equation}
    \boldsymbol{\Gamma}(\theta) = \begin{bmatrix}
        T_p & 0 & 0 \\
        0 & T_s & 0 \\
        0 & 0 & 1
    \end{bmatrix}
\end{equation}

where $T_p$ and $T_s$ is the transmission of the lens for p- and s-polarized light respectively. In this paper, the transmission was measured experimentally. The next element is not a matrix, but simply a scalar function. This function, $A(\theta)$, is the apodization function of the lens. This can be found using the Abbe-Sine condition, and is given by:

\begin{equation}
    A(\theta) = \sqrt{\frac{n}{\cos(\theta)}} \text{\quad\quad} A'(\theta) = \sqrt{\frac{\cos(\theta)}{n}}
\end{equation}

where $A(\theta)$ is for a collimating lens, and $A'(\theta)$ is for a focusing lens. The next matrix is $\textbf{R}_\textbf{x}(\alpha)$, which is used to rotate the electric field around the x-axis. In this paper, this matrix is used to simulate the rotation of the optical axis between O2 and O3. This matrix is given by:

\begin{equation}
    \textbf{R}_\textbf{x}(\alpha) = \begin{bmatrix}
        1 & 0 & 0 \\
        0 & \cos(\alpha) & -\sin(\alpha) \\
        0 & \sin(\alpha) & \cos(\alpha)
    \end{bmatrix}
\end{equation}

where $\alpha$ is the tilt of the optical axis. The next matrix is $\textbf{R}_\textbf{y}(\theta)$, which is used to rotate the electric field around the y-axis. This is used to map the electric field from the meridional and sagittal planes into s- and p-polarized light. This matrix is given by:

\begin{equation}
    \textbf{R}_\textbf{y}(\theta) = \begin{bmatrix}
        \cos(\theta) & 0 & -\sin(\theta) \\
        0 & 1 & 0 \\
        \sin(\theta) & 0 & \cos(\theta)
    \end{bmatrix}
\end{equation}

where $\theta$ is the angle of the rays relative to the optical axis. 

The last matrix is $\textbf{F}_\textbf{T}$, which is a transmission matrix for a refractive index change. This matrix is identical in form to $\boldsymbol{\Gamma}(\theta)$. However, the transmissions for this matrix are calculated using the Fresnel equations. For our case, where the transmission through the refractive index change was measured along with the O3 transmission, this matrix was substituded with an identity matrix of equal size.

\section*{Sampling criterion}

For the simulations to be valid, the electric field traced through the system needs to be properly sampled. For this to be true we need to sample fine enough for the phase to not change by more than $\pi$ between two pixels. For this to be true, we can use the sampling condition outlined by Leutenegger et al.:

\begin{equation}
    N_m > \frac{2\text{NA}_f^2}{\sqrt{n_f^2-\text{NA}_f^2}}\frac{|z|}{\lambda_0}
\end{equation}

where $N_m$ is the minimum needed sampling points, $n_f$ is the refractive index in the image space, $z$ is the maximum axial distance from the focal point, $\text{NA}_f$ is the numerical aperture of the focusing lens, and $\lambda_0$ is the wavelength in vacuum. In our simulations, the sampling volume is a cube with isotropic resolution in the sample space. This means our $|z|$ can be found using:

\begin{equation}
    |z| = \frac{v N_c M_a}{M_t}
\end{equation}

where $v$ is the voxel size, $N_c$ is the number of voxels in the axial direction, $M_a$ is the axial magnification of the system, and $M_t$ is the transverse magnification of the system. Assuming a silicon immersion $O_1$, dry $O_2$, and a snouty $O_3$, we get $M_t=56$ and $M_a=2240$. Using a matched 0.025NA $T_3$ we get:

\begin{equation}
    N_m \gtrsim 0.05 \frac{v}{\lambda_0} N_c
\end{equation}

We know that $N_m$ must exist on the interval $N_m\in [0,N_c]$. As we will always sample with the full array of simulated voxels, we can set $N_m$ to its maximum value $N_c$, leading to the condition $v \lesssim 20\lambda_0$. Assuming the sample should be Nyquist sampled, we also get the condition:

\begin{equation}
    v \leq n_v < \frac{\lambda_0}{2 \text{NA}_{O_1}}\frac{M_t}{2}
\end{equation}

where $n_v$ is the Nyquist voxel. Assuming a high NA $O_1$ (NA$\in(1,1.4)$), we get the condition:

\begin{equation}
    v < \frac{\lambda_0 M_t}{4 \text{NA}_{O_1}} < 10\lambda_0
\end{equation}

Assuming Nyquist sampling, we can then conclude that our sampling rate is sufficient to avoid phase changes of more than $\pi$. The voxel size can also be determined using the equation outlined in\cite{leutenegger2006fast}:

\begin{equation}
    v = \frac{N_c \lambda_0}{N_p \text{NA}_f}
\end{equation}

where $N_p$ is the total number of pixels in one axis after padding the image to avoid aliasing. Redefining $N_p/2N_c = s$ where s is the scaling factor of the padding. To avoid aliasing, the scaling factor should be at least 2\cite{leutenegger2006fast}. This results in the condition:

\begin{equation}
    v = \frac{\lambda_0}{2 s \text{NA}_f} < 10\lambda_0
\end{equation}

This means the requirement for our voxel size will always be met for a suitable scaling factor.

\section*{OTF background}

To find the extent of an OTF, we need to find where the crossing from signal to backgound happens. To find this cutoff power, we use the assumed composition of our signal:

\begin{align}
    X &= P(C_p) + G(\sigma_g) + b \\
    X &= C_p + \eta_p + \eta_g + b
    \label{eq:signal}
\end{align}

where $P$ is a poisson distribution of the photon count $C_p$ resulting in noise $\eta_p$, $G$ is gaussian noise of $\sigma_g$ resulting in noise $\eta_g$, and $b$ is a bias offset. The noise $\eta_p$ and $\eta_g$ are assumed to be independent and spectrally white. We can then find the variance of the signal using:

\begin{equation}
    Var(X_i) = \cancelto{0}{Var(C_{p,i})} + Var(\eta_{p,i}) + Var(\eta_{g,i}) + \cancelto{0}{Var(b_i)} = \sigma_{p,i}^2 + \sigma_{g,i}^2
\end{equation}

where $i$ is the pixel index. As the noise is spectrally white the variance for each pixel is independent. The variance of the entire signal is then given by:

\begin{equation}
    Var(X) = \frac{\sum_i Var(X_i) + \cancelto{0}{\sum_{i\neq j}Cov(X_i,X_j)}}{N_c^3} = \overline{\sigma_p^2} + \overline{\sigma_g^2}
\end{equation}

where $\overline{\sigma_p^2}$ and $\overline{\sigma_g^2}$ are the average variance of the poisson and gaussian noise respectively. We know that the variance of the poisson noise is equal to the mean, and the variance of the gaussian noise is given by the camera manufacturer by the electron RMS value. This means we can find the variance of the signal:

\begin{equation}
    Var(X) = \overline{C_p} + \sigma_{RMS}^2
\end{equation}

where $\overline{C_p}$ is the average signal power without the bias offset. To correlate this variance to the OTF, we assume the signal can be decomposed as in equation \ref{eq:signal}. As the Fourier transform is a linear operation, we can then assume the OTF can be decomposed as:

\begin{equation}
    \hat{X} = \hat{C_p} + \hat{\eta_p} + \hat{\eta_g} + \hat{b}
\end{equation}

where $\hat{X}$ is the Fourier transform of $X$. We then assume the signal ground truth transforms into a perfect OTF with no background and DC term equal $\overline{C_p}$. The bias offset will only have a frequency component in the DC term equal to $b$. As the noise terms are spectrally white, they will have a constant power across all frequencies. This power can be found using Wiener-Khinchin theorem\cite{stein2000computer}:

\begin{equation}
    \overline{S}_X = |\hat{X}|^2
\end{equation}

where $\overline{S}_X$ is the average power spectral density of $X$. We can find the average power of the Fourier transform of the noise terms:

\begin{equation}
    \overline{S}_{\eta} = \sum_{h} \gamma_{eta}(h)e^{-2\pi i \nu h}
\end{equation}

where $\gamma_{eta}(h)$ is the autocovariance of the noise term $\eta$, and $\nu$ is the frequency. As the noise is spectrally white, this only leads to the variance of the noise, and we get:

\begin{align}
    \overline{S}_{\eta} &= \overline{S}_{\eta_p} + \overline{S}_{\eta_g} = C_p + \sigma_{RMS}^2 \\
    |\hat{\eta}| &= \sqrt{\overline{C_p} + \sigma_{RMS}^2}
\end{align}

as we don't know $C_p$, we can estimate it using the DC term of the OTF. As the DC term is composed of the signal and the bias offset, we get:

\begin{equation}
    \overline{C_p} = DC_{\hat{X}}-b 
\end{equation}

where $DC_{\hat{X}}$ is the DC term of the OTF. Using this we can find the average spectral power of the background as:

\begin{equation}
    |\hat{\eta}| = \sqrt{DC_{\hat{X}} - b + \sigma_{RMS}^2}
\end{equation}

By subtracting this background level from the OTF, we define the cutoff frequency as the first zero-crossing of the resulting OTF. Then, assuming the OTF is an ellipsoid, we can find the PSF volume and are from the resulting resolution limits.

\printbibliography